\documentstyle[pra,aps,epsf,graphicx]{revtex}

\newcommand{\bge}{\begin{equation}}
\newcommand{\ee}{\end{equation}}
\newcommand{\bgc}{\begin{center}}
\newcommand{\ec}{\end{center}}
\newcommand{\bgea}{\begin{eqnarray}}
\newcommand{\eea}{\end{eqnarray}}
\newcommand{\bgeas}{\begin{eqnarray*}}
\newcommand{\eeas}{\end{eqnarray*}}

\newcommand{\bnabla}{{\bf \nabla}}
\newcommand{\hcc}{H_{\mbox{\scriptsize cc}}}
\newcommand{\hcd}{H_{\mbox{\scriptsize cd}}}
\newcommand{\hdc}{H_{\mbox{\scriptsize dc}}}
\newcommand{\hdd}{H_{\mbox{\scriptsize dd}}}

\begin{document}

\title{
  \begin{flushleft}
    {\footnotesize BU-CCS-981001}\\[1.0cm]
  \end{flushleft}
  \bf Lattice-Gas Simulations of Ternary Amphiphilic Fluid Flow in
  Porous Media~\footnote{To appear in {\it International Journal of
  Modern Physics C} (1998)}}
\author{
  P.V. Coveney, J.-B. Maillet and J.L. Wilson\\
  {\footnotesize Schlumberger Cambridge Research,}\\
  {\footnotesize High Cross, Madingley Road, Cambridge CB3 OE5, U.K.}\\
  {\footnotesize \tt \{coveney,maillet\}@cambridge.scr.slb.com}\\[0.3cm]
  P.W. Fowler\\
  {\footnotesize Cavendish Laboratory, University of Cambridge,}\\
  {\footnotesize Madingley Road, Cambridge CB3 0EL, UK}\\[0.3cm]
  O. Al-Mushadani\\
  {\footnotesize Department of Physics, University of Oxford,}\\
  {\footnotesize 1 Keble Road, Oxford OX1 3NP, UK}\\[0.3cm]
  B.M. Boghosian\\
  {\footnotesize Center for Computational Science, Boston University,}\\
  {\footnotesize 3 Cummington Street, Boston, Massachusetts 02215, U.S.A.}\\
  {\footnotesize{\tt bruceb@bu.edu}}\\[0.3cm]
  }
\date{\today}
\maketitle

\begin{abstract}
We develop our existing two-dimensional lattice gas model to simulate
the flow of single phase, binary immiscible and ternary amphiphilic
fluids. This involves the inclusion of fixed obstacles on the lattice,
together with the inclusion of ``no-slip'' boundary conditions. Here we
report on preliminary applications of this model to the flow of such
fluids within model porous media. We also construct fluid invasion
boundary conditions, and the effects of invading aqueous solutions of
surfactant on oil-saturated rock during imbibition and drainage are
described.
\end{abstract}

\section{Introduction}

The lattice gas automaton (LGA) model introduced by Boghosian, Coveney
and Emerton\cite{bib:bce} has been used to investigate a variety of
amphiphilic phenomena including the growth kinetics of binary immiscible
fluid and ternary microemulsion systems\cite{bib:ecb,bib:wcb}, the
effect of shear on ternary systems\cite{bib:ewcb} and self-reproducing
micelles\cite{bib:ceb}. In this article we describe the developments
that have been made to allow invasive flow within a porous medium to be
studied using this model. These developments open up a large area to
possible investigation and we shall only give a very brief overview of
the work done to date. We emphasise that the results presented here are
preliminary and that considerably more work is needed in this area; much
of this is currently in progress.

\section{Development of the model}

For a discussion of the original model, the reader is referred to
Boghosian, Coveney and Emerton\cite{bib:bce}. In the general case, the
model admits the presence of three different species of particle, which
are distinguished by their ``colors'' (red for oil, blue for water, and
green for surfactant). The modifications to the model which are made in
this paper fall into two categories: the implementation of bulk flow and
the alteration of boundary conditions.

The study of bulk flow had previously been limited to the case of linear
shear flow\cite{bib:ewcb}. The new techniques we have implemented use a
similar method to that of Olson and Rothman\cite{bib:ols1,bib:ols2}:
momentum is added at random sites throughout the fluid. As the particles
are situated on a discrete triangular (FHP) lattice with six lattice
vectors, the change in momentum at any given site may be quite large
although the global effect is small. This is not unreasonable as the
propagation and collision steps of the model (which impart viscosity to
the fluids) distribute this momentum in a relatively short
time. However, while Olson and Rothman manipulate each particle
separately, in our code use is made of the pre-existing look-up tables
for the maximum and minimum momentum in the y-direction (vertical) that
were used for shear flow\cite{bib:ewcb}.  Previously, sites on opposite
sides of the lattice were looped over in a stochastic manner and their
momentum updated until the shear boundary condition was met. For bulk
flow the {\em entire} lattice is looped over and either:

\begin{enumerate}

\item enough sites are updated for the total average momentum of the
particles to be equal to a set value; this is called the implementation
of a `pressure gradient'. This is not a true pressure gradient, since
pressure is proportional to the density of the fluid. This is the flow
implementation used throughout this article;

\end{enumerate}

\noindent or

\begin{enumerate}

\item enough sites are updated so that the total additional momentum
applied to the system of particles in each timestep is equal to a
constant, thus simulating bulk flow under gravity, i.e. \emph{buoyancy}
effects.

\end{enumerate}

To model flow through porous rock, a channel or any other obstacle
structure, a no-slip boundary condition needs to be set up at the
interface between the particles and a general obstacle matrix. The
no-slip boundary condition applied to a Navier-Stokes fluid corresponds
to a `bounce-back' step in the lattice-gas model\cite{bib:rz,bib:rbook}.

There are three components to the successful implementation of a
realistic no-slip boundary condition\cite{bib:rz}:

\begin{enumerate}

\item First, in the initial set-up of the lattice, the obstacle matrix
has to be cleared of all particles.  It will then remain so for all
time.

\item Second, a simple ``bounce-back'' operation is carried out, in
which a particle colliding with an obstacle has its momentum reversed in
sign.

\item Finally, obstacles may be given a specified color charge, thus
assigning certain wettabilities to the simulated rock species.

\end{enumerate}

In the present paper, which presents purely preliminary results, only
the effects of an obstacle site color density of $\pm 7$ are reported
(equivalent to that for a site \emph{full} of either red or blue
particles), but it is possible to alter the `strength' of each site's
wettability to any integer value between these values.

To allow invasive flows to be studied, the model maintains horizontal
periodic boundary conditions, but is no longer periodic in the vertical
(forced) direction. We must concern ourselves with particles that stray
off the top and bottom of the lattice. Conserving the mass of the
particles on the lattice imposes restrictions on the fate of these
particles: we cannot, for example, arbitrarily add particles to the base
of the lattice since this violates the conservation of mass.  Hence we
are forced to wrap the particles from top to bottom and {\em vice
versa}, but if an \emph{oil} particle propagates off the top of the
lattice, then on the subsequent timestep we re-introduce this particle
as a \emph{water} particle at the base of the lattice moving in the same
direction as when it left the lattice. This satisfies the mass
conservation law and also imitates the physical process under study. An
extension of this is required if surfactant is present in the invading
phase.

A further modification is required to convince the particles on the top
and bottom rows of the lattice that they are below and above infinite
columns of oil and water respectively. This is done by adding two extra
``invisible'' rows which affect the collisions on the top and bottom
rows, but which play no part in particle propagation.

Note that in order to incorporate the most general form of interaction
energy within the model system, a set of coupling constants $\alpha,
\mu, \epsilon, \zeta$, are introduced, in terms of which the total
interaction energy can be written as
\bge
\Delta H_{\mbox{\scriptsize int}}
       =   \alpha \Delta \hcc +
         \mu \Delta \hcd +
         \epsilon \Delta \hdc +
         \zeta \Delta \hdd.
\label{eq:tiw}
\ee
These terms correspond, respectively, to the relative immiscibility of
oil and water, the tendency of surrounding dipoles to bend round oil or
water particles and clusters, the propensity of surfactant molecules to
align across oil-water interfaces and a contribution from pairwise
(alignment) interactions between surfactants. In this report, as in
previous studies\cite{bib:bce,bib:ecb,bib:wcb,bib:ewcb} $\beta = 1.0$
and the coupling constants are given the values
\bgeas
  \alpha &=& 1.0  \\
     \mu &=& 0.001\\
\epsilon &=& 8.0  \\
   \zeta &=& 0.005,
\eeas
strongly encouraging surfactant molecules to accumulate at oil-water
interfaces while maintaining the normal oil-water immiscible behavior.\\
Different porous media have been constructed. The random location of
obstacles (spheres or squares) is the method most commonly used to
construct an artificial porous medium.  In the present work, we have
used a novel modification of this technique which allows one to control
the size and connectivity of the obstacle matrix without introducing
artificial symmetries. In this approach, an initial simulation is run
leading to a microemulsion phase (droplet or sponge).  Droplets of, say,
oil are formed and grow with time. When their size reaches some desired
value, the positions of these oil particles are stored as obstacle sites
for later simulations. Suitable modification of the initial conditions
(reduced densities, flow conditions and so on) produces a wide range of
porous media for subsequent study.

\section{Darcy's Law}

\begin{figure}[htbp]
\vspace*{13pt}
\begin{center}
\rotatebox{270}{\includegraphics{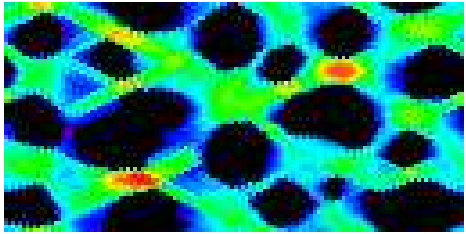}}
\end{center}
\caption{The porous medium used in the
simulation discussed in section 3, together with a coloured velocity
profile for a single component fluid flow.}
\label{porousmed}
\end{figure}

Darcy's Law relates the flux of a \emph{single phase} fluid through a
porous medium to the gradient of pressure across the
medium. Mathematically, Darcy's law can be written:
\begin{equation}
u = -\frac{k}{\mu} X
\label{eq:dar1}
\end{equation}
\noindent where the constant {\it k} is the {\it permeability} of the
medium, {\it $\mu$} is the dynamic viscosity and {\it $X= -\bnabla P
+\rho g$} is the fluid forcing (pressure gradient plus gravitational
force density). This relation was verified using our code on different
porous media. The results presented in this section are obtained using
the porous medium shown in Fig.~\ref{porousmed}.

\subsection{Binary Immiscible Flow}

There is no widely accepted extension of Eq.~(\ref{eq:dar1}) which deals
with the case of two-phase or multiphase flows within porous media. It
is frequently asserted that Eq.~(\ref{eq:dar1}) generalises to two or
more similar but completely uncoupled equations, in which the {\it
relative permeabilities} make an appearance.
Experimental\cite{bib:kalaydjian} and
numerical\cite{bib:rothman,bib:gunstensen} work has suggested an
extension of Eq.~(\ref{eq:dar1}) including coupling terms between the
two fluids:

\begin{equation}
 \left( \begin{array}{c}
		u_w \\
		u_n \\
		\end{array} \right) 
=k \left( \begin{array}{cc}
		\kappa_{ww} & \kappa_{wn} \\
		\kappa_{nw} & \kappa_{nn} \\
		\end{array} \right) 
 \left( \begin{array}{cc}
		F_w \\
		F_n \\
		\end{array} \right), 
\label{eq:dar2}
\end{equation}
\noindent where the subscript `w' indicates a `wetting' phase and the
subscript `n' indicates a `non-wetting' phase. The new transport
coefficients {\it $\kappa_{ij}$} are {\it relative permeabilities}
(diagonal terms) and coupling coefficients (off-diagonal terms). In
general these {\it $\kappa_{ij}$} will depend on the viscosity of each
fluid and the composition of the mixture. The force acting on particles
in Eq.~(\ref{eq:dar1}) has been replaced by the gravitational force, as
used in the previously described simulations.  This technique is closely
related to several other implementation methods\cite{bib:ols1,bib:ols2}.

Much has been made of the reciprocity of the {\it $\kappa_{wn}$} and
{\it $\kappa_{nw}$} coupling coefficients, and its possible relationship
with Onsager's reciprocity relations in linear non-equilibrium
thermodynamics\cite{bib:ons1,bib:ons2}.  Some theoretical work has been
done on this subject for the linear regime\cite{bib:pf}, but no
explanation exists for the generally nonlinear behavior displayed in
lattice-gas models. There is also a quite surprising lack of
experimental data available by means of which to resolve this issue.

\begin{figure}[htbp]
\vspace*{13pt}
\begin{center}
\includegraphics[width = 10cm]{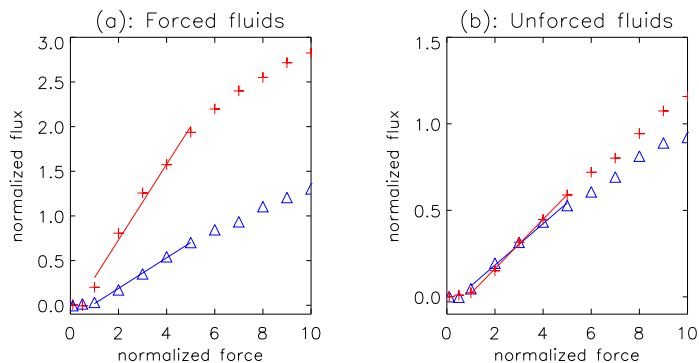} 
\end{center}
\caption{Darcy's law behavior for binary immiscible fluids. Triangles
correspond to the response of water and crosses to the response of
oil. The normalized flux and force are defined in the same way as by
Olson and Rothman, for a forcing of $0.001$ per lattice site. The lines
represent least-squares fits to the data.}
\label{fig:bindarcy}
\end{figure}

Fig.~\ref{fig:bindarcy} shows the results of measurements made on the
flow of wetting and non-wetting phases when either of these phases are
forced in a water wetting porous medium; the graphs describe the
relative permeabilities (seen as the supposedly constant gradients of
the curves) from the matrix of Eq.~(\ref{eq:dar2}) for reduced densities
of oil and water equal to $\rho_{i}=0.25$. Each data point on a plot
represents the result from a single run, averaged over $20,000$ time
steps for a 64$\times$128 lattice.  In this binary immiscible case, the
maximum `pressure' that can be applied to the system is approximately
$0.025$ momentum units per lattice site (the ``gravity'' condition is
used).  The relative permeabilities (diagonal elements of the matrix in
Eq.~(\ref{eq:dar2})) appear to be constant over most `pressures', with a
least-squares fit plotted on the figures.

As in the three-dimensional simulations of Olson and
Rothman\cite{bib:ols1}, when the fluids are forced the curve pertaining
to the non-wetting fluid, Fig.~\ref{fig:bindarcy}(a), shows capillary
effects at low forcing levels. Droplets of the non-wetting fluid are
trapped in the medium, resulting in a disconnected phase. Nevertheless,
the wetting fluid also exibits a nonlinear behavior in the same
regime. This is because, in the $2D$ porous medium, the obstacle matrix
is not continuous so a certain `pressure' is required before the wetting
fluid can break through its surface tension and flow. However, the
non-wetting fluid is `lubricated' by the wetting fluid, causing it to
flow readily.  This `lubrication' is mainly responsible for the
different slopes, or different relative permeabilities, for the two
phases.  The differences in the response of water and oil when they are
forced arise directly from the wettability of the rock versus water.

From inspection of Fig.~\ref{fig:bindarcy}(b) it is clear that the
definition of the coupling coefficients as constants is approximately
valid, for pressure gradients above a capillary threshold. Moreover, it
can be seen that the responses of the unforced fluids are roughly
symmetric. The results are thus consistent with Onsager reciprocity.  At
high `pressures' the water `slides' up around the obstacles and around
the oil drops, imposing a slowly increasing force on the non-wetting
phase.

It is important to notice the various scales on the flow axes: the
coupling of the two phases is significant. This coupling is the
consequence of the large amount of fluid-fluid interface, and is related
to the average pore size in the porous medium.  Taking the relative
permeabilities to be constant over all `pressures' above the capillary
threshold, their calculated values are ${\it \kappa_{ww}} = 0.17\pm
0.0007$ and ${\it \kappa_{nn}} = 0.42\pm 0.003$.  The values for the
coupling coefficients are ${\it \kappa_{wn}} = 0.14\pm 0.0009$ and ${\it
\kappa_{nw}} = 0.12\pm 0.0003$.

\subsection{Ternary Amphiphilic Flow}

Fig.~\ref{fig:terdarcy} shows the results of measurements made on the
flow of wetting and non-wetting phases when either of these phases are
forced (in the same water-wetting porous medium).  The reduced densities
of wetting and non-wetting fluids was $0.2$ and the simulations were
performed at the same total density as before.  These results are
qualitatively similar to those obtained in the binary immiscible
case. However, some differences do exist:

\begin{figure}[htbp]
\vspace*{13pt}
\begin{center}
\includegraphics[width = 10cm]{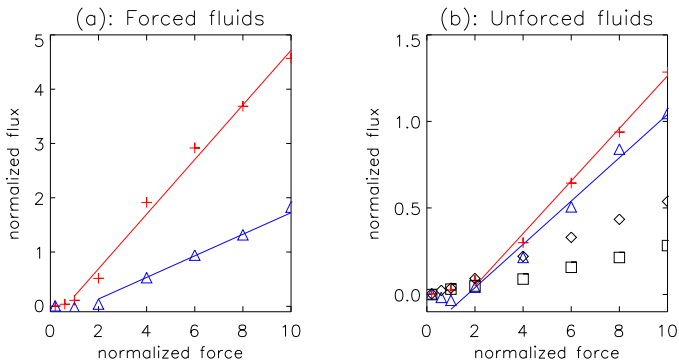} 
\end{center}
\caption{Darcy's law behavior of ternary amphiphilic fluids. Triangles
correspond to the response of water and crosses to the response of
oil. Squares and diamonds correspond to the response of surfactant when
oil or water is forced respectively.  The normalized flux and force are
defined in the same way as by Olson and Rothman, for a forcing of
$0.0005$ per lattice site.}
\label{fig:terdarcy}
\end{figure}

\begin{enumerate}
\item Firstly the values for the relative permeabilities in the ternary
case are ${\it \kappa_{ww}} = 0.21\pm 0.005$ and ${\it \kappa_{nn}} =
0.51\pm 0.084$.  The values for the coupling coefficients are ${\it
\kappa_{wn}} = 0.15\pm 0.004$ and ${\it \kappa_{nw}} = 0.12\pm 0.005$.
The marked change in ${\it \kappa_{ww}}$ and ${\it \kappa_{nn}}$ is due
to a lowering in the surface tension between the oil and water phases
now that surfactant is present at the interface; the water then flows
more easily through the large channels within which the non-wetting oil
would otherwise sit.
\item The flux of the fluids when they are not forced is roughly the
same as in the binary case. Their responses are still essentially
symmetric.  The response of surfactant is also shown, which varies
linearly with the forcing level.
\item The capillary threshold is lower than in the binary case (here
equal to $0.0005$). This can be explained by the reduction in surface
tension when surfactant is present (oil can pass more easily through
narrow channels).
\end{enumerate}

\section{Study of the effect of surfactant on fluid invasion in porous media}

No lattice gas automaton model has previously investigated the effect of
surfactant on an oil/water system undergoing invasive flow. We start
from a porous medium filled with oil into which water invades; this
invading phase may contain varying amounts of surfactant and the effect
of this on oil production is studied qualitatively in this preliminary
study. The obstacle matrix may also be given different wetting
properties (oil-, water- or non-wetting). With water as the invading
phase, the case with oil-wetting obstacles corresponds to drainage and
that with water-wetting obstacles to imbibition.  In the present model,
only particles on lattice sites containing water are forced.  Unlike
previous non-invasive studies, the number of forcing particles on the
lattice increases with time. Although this is taken into account in
deciding the maximum forcing amplitude that can be applied, it can lead
to spurious effects. When there is an insufficient quantity of water
present, the water is ``over-driven'' and, for example, small droplets
of water can detach from the bulk phase and are hence propelled into the
oil.  Moreover, the flow of surfactant particles is entirely dependent
on the transfer of momentum from forced water particles. Thus, although
a specified proportion of surfactant is placed on the bottom row of the
lattice, some may fail to propagate into the bulk of the invading
phase. Both of these features limit the preliminary study reported here
to being a qualitative approach only. We plan to return in the future
with a more quantitative study.
\begin{figure}[htbp]
\vspace*{13pt}
\begin{center} 
\includegraphics[width = 10cm]{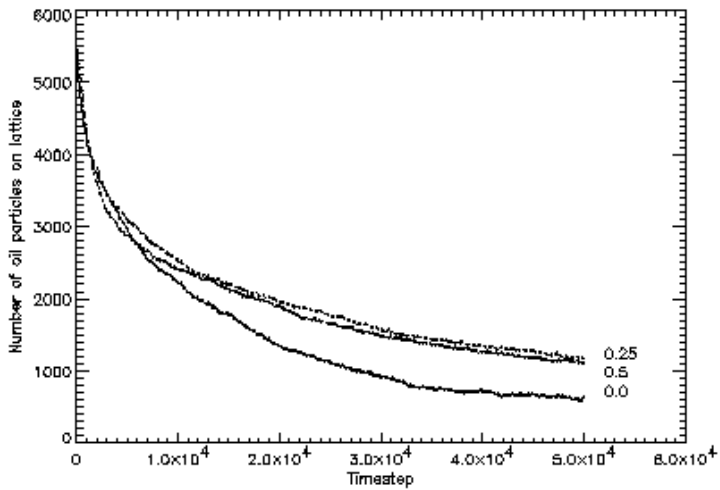} 
\end{center}
\vspace*{13pt}
\caption{Graph showing how the amount of oil on the lattice changes
with time for drainage with different relative concentrations of
surfactant{\footnotesize\it}.}
\label{fig:draingraph}
\end{figure}

The effect of aqueous surfactant concentration on oil production is
investigated by tracking the amount of oil on the lattice as a function
of time. We first consider the phenomenon of drainage
(Fig.~\ref{fig:draingraph}).  More oil is expelled off the lattice if
\emph{no} surfactant is present.  The difference between concentrations
0.25 and 0.5 is too small to allow any conclusions about their relative
magnitudes to be drawn. A second study revealed a similar effect. (Both
studies used a 64$\times$64 lattice and a forcing amplitude of $0.010$.)

\begin{figure}[htb]
\vspace*{13pt}
\begin{center} 
\includegraphics[width = 10cm]{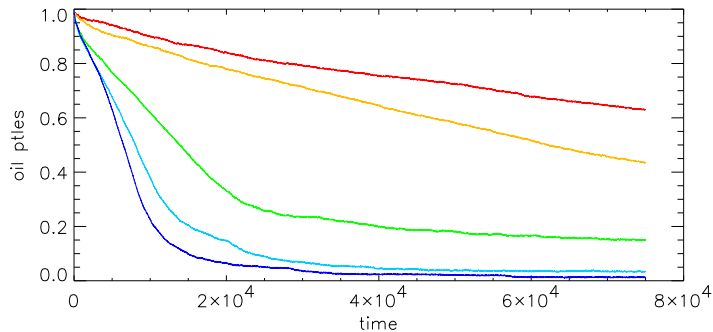}	
\vspace*{13pt}
\end{center}
\caption{Effect of the applied force in imbibition simulations. The
different forcing levels (from top to bottom) are 0.0005, 0.001, 0.005,
0.01, 0.025}
\label{fig:percgraph}
\end{figure}

Now let us turn to the case of imbibition. The two imbibition
investigations which we have performed in these preliminary
investigations show quite marked variations. In one study, we found that
the presence of surfactant improved the recovery of oil
dramatically. However, a second study found that the presence of
surfactant made little difference to this process. Each investigation
used a different obstacle matrix in addition to other differences in the
parameters used, all of which no doubt contribute to produce these
differences.  We believe that, although the behavior we have found so
far is inconclusive, further more detailed investigations will produce
interesting results from which more systematic conclusions may be drawn,
and we plan to publish our findings in due course.

One expects that percolation plays a large role in the behavior of these
systems\cite{bib:wilkinson}.  Fig.~\ref{fig:percgraph} shows imbibition
simulations for different driving forces.  One can see that as the
driving force increases, not only the time needed to reach the
asymptotic behavior decreases but so does the residual oil saturation.
Prior to percolation, the invading fluid (water) phase bodily drives oil
off the lattice. The variation of the number of oil particles versus
time is roughly linear.  After the percolation threshold for water, the
decrease of the number of oil particles is less pronounced. The end of
oil percolation leads to the asymptotic regime (defining the residual
oil saturation).  This overall behavior depends on the driving force,
the concentration of surfactant present (and obviously on the porous
medium). For example, it is possible to envisage a scenario where the
percolation channels are wide enough to sustain the flow of the invading
phase and hence no further oil is driven from the lattice, leading to
high residual oil saturation. The influence of percolation on this
behavior is currently under investigation.

\begin{figure}[htbp]
\vspace*{13pt}
\begin{center} 
\includegraphics[width = 10cm]{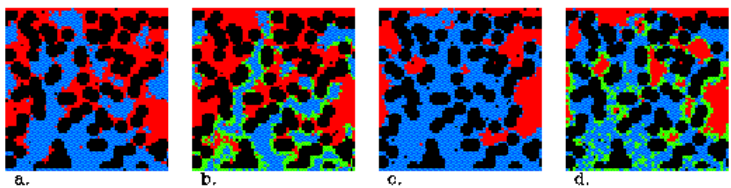} 
\end{center} 
\caption{Invasive flow of water in drainage and imbibition.  The
snapshots show different configurations at timestep 8,000: (a), (b)
drainage without and with surfactant, respectively; (c),(d) imbibition
without and with surfactant, respectively{\footnotesize\it}.}
\label{fig:invadesnapshots}
\end{figure}

Fig.~\ref{fig:invadesnapshots} shows four graphical snapshots of the
system at the same timestep but with different configurations.
Figs.~\ref{fig:invadesnapshots} (a) and~\ref{fig:invadesnapshots} (b)
show drainage without and with surfactant present in the invading phase
respectively.  Figs.~\ref{fig:invadesnapshots} (c)
and~\ref{fig:invadesnapshots} (d) are similar but show the case of
imbibition. All are from timestep 8,000.  These snapshots are intended
to illustrate qualitative differences only.

Figs.~\ref{fig:invadesnapshots} (a) and~\ref{fig:invadesnapshots} (b)
suggest that the presence of surfactant permits the formation of stable
percolating channels by the invading aqueous phase, leading to greater
oil retention within the porous medium. Figs.~\ref{fig:invadesnapshots}
(c) and~\ref{fig:invadesnapshots} (d), which are taken from the first of
the two studies of imbibition mentioned above, suggest that, at least if
there are relatively large voids in the porous medium, aqueous
surfactant can break down and hence remove--by emulsification and/or
micellization--the pockets of oil trapped within the rock. Nevertheless,
the time scale over which such phenomena occur could slow down the oil
production process.  We emphasise again, however, that these are purely
preliminary results valid in specific cases. Further work will be
necessary before it is possible to establish whether any general
conclusions can be drawn.

\section{Discussion}

We have described an extension of our hydrodynamic lattice-gas model to
handle invasive flows within two-dimensional porous media.  This has
allowed us to investigate a generalization of Darcy's law for multiphase
flows, including effects due to the presence of surfactant.  Invasion
studies in porous media produced some interesting preliminary results,
particularly in connection with the presence of surfactant within the
invading phase. However, to permit a more complete and quantitative
analysis, further developments and more detailed simulation studies are
required; indeed, some of these are already underway.  For example, one
might consider devising alternative fluid forcing algorithms for
improved handling of the flow of the invading phase; {\em inter alia},
as noted above, surfactant and water could be forced together. A more
detailed investigation of the role of percolation is required.  The
reliability of our results would be enhanced by the use of larger
lattices and increased statistical averaging.  Throughout the work
discussed here, both water-wetting and oil-wetting porous rock obstacles
have been made maximally wetting. If the color charge on the obstacle
sites were reduced, the phenomenological behavior could be expected to
change significantly.  These comments reflect the quandary faced when
deciding what to investigate with this model. Considering merely
invasive studies, there are evidently many important areas now open for
investigation and many parameters which could be altered, although this
report has only touched on a few of these. Not of least importance is
the structure of the porous medium itself: one can expect significant
changes in flow behavior in going from two to three dimensions. We hope
to address more closely the real world using a three-dimensional version
of our model\cite{bib:bc} at some stage in the future.  However,
two-dimensional experimental systems exist for studying porous media
flows--so-called ``micromodels''--and we believe that there is already
scope for a valuable dialogue between our simulations and experiments in
this area\cite{bib:maloy}.

\section{Conclusion}

The results on binary immiscible and ternary amphiphilic fluids show
that the presence of surfactant, when either phase is forced, increases
the flow of the forced phase. The relative permeabilities seem well
behaved in both the binary immiscible and ternary amphiphilic fluids,
while the coupling coefficients are more dependent on the actual
geometry and dimensionality of the obstacle matrix.  Moreover, the
capillary threshold is lowered by the incorporation of amphiphilic
particles.

The study of ternary amphiphilic invasion into a porous medium showed
that the introduction of surfactant into the invading phase alters
drainage and imbibition. Preliminary results have shown that the
addition of surfactant seems to impair drainage. However, we cannot yet
make any conclusive remarks about the effect of surfactant on
imbibition.

\section*{Acknowledgements}

The travel required for this collaboration was supported in part by NATO
grant number CRG950356.  BMB was supported in part by the United States
Air Force Office of Scientific Research under grant number
F49620-95-1-0285, and in part by AFRL.

\end{document}